\documentclass[10pt,prb,aps,twocolumn,showpacs,amsmath,amssymb,superscriptaddress]{revtex4}
\usepackage{graphicx}
\usepackage{dcolumn}
\usepackage{bm}
\usepackage{epsfig,psfrag,amsmath,amssymb,float}
\input{epsf}
\begin{document}

\title{Effect of an Electron-phonon Interaction on the One-electron\\
Spectral Weight of a d-wave Superconductor}

\author{A.~W.~Sandvik}
\email{asandvik@abo.fi}
\affiliation{Department of Physics,{\AA}bo Akademi University,
Porthansgatan 3, FIN-20500 Turku, Finland}
\affiliation{Department of Physics, University of California, Santa
Barbara, CA 93106-9530}

\author{D.~J.~Scalapino}
\email{djs@vulcan2.physics.ucsb.edu}
\affiliation{Department of Physics, University of California, Santa
Barbara, CA 93106-9530}

\author{N.E.~Bickers}
\email{bickers@physics.usc.edu}
\affiliation{Department of Physics, University of Southern California,
Los Angeles, CA 90089-0484}

\date{\today}

\begin{abstract}

Here we analyze the effects of an electron-phonon interaction on the
one-electron spectral weight $A(k, \omega)$ of a $d_{x^2-y^2}$
superconductor. We study the case of an Einstein phonon mode with various
momentum-dependent electron-phonon couplings and compare the structure
produced in $A(k, \omega)$ with that obtained from coupling to the magnetic
$\pi$-resonant mode. We find that if the strength of the interactions are
adjusted to give the same renormalization at the nodal point,
the differences in $A(k, \omega)$ are generally small but possibly 
observable near $k=(\pi,0)$.

\end{abstract}

\pacs{79.60, 74.72.-h, 74.25.Kc}

\maketitle

\section{Introduction}

The role of the electron-phonon coupling in the high $T_c$ cuprates remains
a puzzle. The initial finding of the absence of a phonon signature in the
temperature dependence of the resistivity\cite{Mar90}
 and the small size of the isotope
effect in the optimally doped cuprates\cite{Fra94}
 suggested that the electron-phonon
interaction played a relatively unimportant role in these
strongly-correlated materials.  However, large isotope effects away from
optimal doping,\cite{Fra94,Cra90} significant phonon renormalization 
induced in the
superconducting state,\cite{BHI00,MRS92,Mcq01,Pyk93} 
and recent interpretations of ARPES data\cite{Lan01,SLIN02}
continue to raise questions
regarding the nature and role of the electron-phonon interaction in the
high $T_c$ cuprates.
 
One point of view is that the effects
of the strong Coulomb interaction act to suppress the electron-phonon
interaction and that while the electron-lattice interaction enters the
problem, it does so on a secondary level coming along as it were for the
ride. For example, in this view the large isotope effects observed in some
of the cuprates away from optimal doping arises from the influence of the
lattice on stripe fluctuations, acting to stabilize these and thus
suppressing superconductivity.\cite{CEKO02} Similarly, the 
superconductivity-induced
phonon renormalization and the possible Englesberg-Schrieffer\cite{ES63}
 signature in
the ARPES data could be interpreted as naturally occurring in an interacting
system but having little effect on the underlying superconducting pairing
mechanism.  Alternatively one might interpret the isotope effect and the
phonon renormalization as supporting the existence of a significant
electron-phonon coupling.  Furthermore, ARPES measurements have been
specifically interpreted in terms of phonon modes that could drive
$d_{x^2-y^2}$ pairing.\cite{SLIN02} Here, we analyze a simple model of an
electron-phonon interaction with the goal of obtaining insight into what
one expects to see in the ARPES data of a $d_{x^2-y^2}$ superconductor with
electron-phonon interactions.

Continuing technological advances along with improved sample quality have
allowed angle-resolved photoemission spectroscopy (ARPES) to probe details
of the energy and momentum structure of the one-electron excitations in the
cuprate materials.\cite{DHS03} Although simplified, the sudden 
approximation leads
to a useful picture in which the ARPES intensity is equal to the square of
a matrix element which depends upon the photon energy, polarization, and
the sample geometry times a product of the single particle spectral weight
\begin{equation}
A(k, \omega) =  \frac{1}{\pi}\ {\rm Im} \lbrace G(k, \omega)\rbrace
\label{one}
\end{equation} 
and a Fermi factor $f(\omega)$. Here $G(k, \omega)$ is the one-electron
Green's function.  Thus, the idea is that from the $k$ and $\omega$
dependence of the ARPES data, one can extract information about the
spectral weight $A(k, \omega)$. Then, from this, one seeks to 
learn about the electron
self-energy $\Sigma(k, \omega)$ and the structure of the effective
interaction. In particular, the role of spin fluctuations and
the $\pi$-resonance on the superconducting state spectral function have
been studied.\cite{Kam01,EN00,EN03} With the recent suggestions\cite{Lan01}
from ARPES measurements that there may be a significant coupling of the
electrons to a phonon with an energy of order 40meV, one would like to
understand how this would effect the ARPES spectrum.

From the number of atoms in a unit cell, it is clear that there are a large
number of phonon modes in the cuprates. Here we will focus on several of
the modes associated with the motion of the O ions. We will treat these
as Einstein phonons. Then for a Hubbard-like model in which the Cu sites
form the Hubbard lattice, the effective electron-electron interaction is 
\begin{equation} 
V(q, \omega) = \frac{2|g(q)|^2\Omega_0}{\omega^2 - \Omega^2_0 + i\delta}.
\label{two}
\end{equation}
If $|g(q)|^2 = |g|^2$ is independent of the momentum transfer, $V(q,
\omega)$ does not couple to the $d_{x^2-y^2}$-pairing channel. This could
model the coupling to the $c$-axis vibration of the apical oxygen.
Alternatively, if the electron-phonon matrix element is momentum
dependent, the interaction given by Eq.~\eqref{two} can couple to the
$d_{x^2-y^2}$-pairing channel.  

The possibility that an electron-phonon interaction could give rise to
$d$-wave pairing has been discussed by various 
authors.\cite{SLIN02,ZK96,SA95,Sca95,ND96,DMFT96,BS96} In one
approach, the $d$-wave pairing interaction occurs as the result of the
interplay of the O half-breathing mode and the exchange 
interaction.\cite{SLIN02} Other
approaches suggest that the Coulomb interaction can lead to a peaking of
the electron-phonon coupling at small momentum transfers which favors
$d_{x^2-y^2}$ pairing.\cite{ZK96,Hua03} This type of momentum dependence 
also occurs directly for certain phonon modes. For example, 
for the Cu-O-Cu buckling-like
mode\cite{SA95,Sca95,ND96,DMFT96,BS96} the square of the electron-phonon
coupling constant is
\begin{equation}
|g(q)|^2 = |g|^2\, \left(\cos^2 \left(\frac{q_x}{2}\right)
 + \cos^2 \left(\frac{q_y}{2}\right)\right).
\label{three}
\end{equation}
Setting $q=k-k^\prime$, the
momentum-dependent part of this coupling factors into a sum of separable
terms
\begin{eqnarray}
&&|g (k-k^\prime)|^2  = \nonumber\\
&&|g|^2 \Bigl(1+\frac{1}{4} (\cos k_x-\cos k_y)
(\cos k^\prime_x - \cos k^\prime_y) + \cdots\Bigr),~~~~~ 
\label{four}
\end{eqnarray}
including additional $(\cos k_x + \cos k_y)$ and $(\sin k_x \pm\ \sin k_y)$
factors. The plus sign in front of the $d$-wave term implies that this type
of phonon exchange provides an attractive channel for $d$-wave pairing.
The key point is that if the electron-phonon coupling $|g (k, k^\prime)|$
falls off at large $|k-k^\prime|$ momentum transfers, then such a phonon
exchange can mediate $d$-wave pairing.

Alternatively, an in-plane O breathing-like mode has 
\begin{equation}
|g(q)|^2 = |g|^2 \left(\sin^2 \left(\frac{q_x}{2}\right) + 
\sin^2 \left(\frac{q_y}{2}\right)\right).
\label{five}
\end{equation}
This increases at large momentum transfers giving rise to a repulsive
interaction in the $d_{x^2-y^2}$-channel. Setting $q=k-k^\prime$ in
Eq.~\eqref{five} one finds that 
\begin{eqnarray}
&&|g(k-k^\prime)|^2 = \nonumber\\
&&|g|^2\Bigl(1-\frac{1}{4}\ (\cos k_x - \cos k_y)
(\cos k^\prime_x - \cos k^\prime_y) + \cdots\Bigr),~~~~~  
\label{six}
\end{eqnarray}
and the minus sign in the second term implies that this phonon suppresses
$d$-wave pairing.\cite{SA95,Sca95,ND96,DMFT96,BS96} 

In Section II we discuss the simplified case of a cylindrical Fermi surface
and a separable phonon mediated interaction. This provides insight into the
differences between the $s$-wave and $d$-wave cases and establishes the
structure of the singularities in the self-energy that are reflected in
$A(k, \omega)$ for an Einstein mode.  While in the actual materials, the
singularities are broadened by the dispersion of the phonon mode,
quasiparticle lifetime effects due to other interactions and impurities, 
as well as
finite temperature effects, these results provide a useful framework for
understanding the structure that appears in $A(k, \omega)$.  

In Section III, we include the effects of a $t-t^\prime$ band structure and
the momentum dependence of the coupling. We consider the three different
electron-phonon coupling constants discussed above and compare these with
the response to the $\pi$-resonance spin fluctuation mode. The analysis of
the $\pi$-resonance mode has been extensively discussed in Refs
\onlinecite{EN00,EN03}. 
Various estimates for the strength of the
$\pi$-resonance-electron coupling have been made.\cite{KKA02,ACENS02}  
Here, we chose this
coupling so that the renormalization of the nodal Fermi velocity is
comparable with that obtained for the phonon coupling. Then we compare and
discuss the spectral weights for the various modes. 
Section IV contains a summary of the results and our conclusions.

\section{A Cylindrical Fermi Surface and an Einstein Phonon}

In this section we consider the case of a cylindrical Fermi
surface and an interaction arising from the exchange of an Einstein phonon
of frequency $\Omega_0$ 
\begin{equation}
V(\theta, \theta^\prime, \omega) = \frac{2|g(\theta, \theta^\prime)|^2
\Omega_0}{\omega^2 - \Omega^2_0 + i\, \delta}.
\label{seven}
\end{equation}
Here, $\theta$ and $\theta^\prime$ denote different $k$ vectors on the
cylindrical Fermi surface. 
With Eq.~\eqref{four} in mind, we will take $|g(\theta, \theta^\prime)|^2$ to
have the separable form
\begin{equation} 
|g(\theta, \theta^\prime)|^2 = |g_z|^2 + |g_\phi|^2 \cos 2\theta \cos
2\theta^\prime.
\label{eight}
\end{equation}
The one-electron Green's function can be written as
\begin{equation}
G(k, \omega) = \frac{Z(\omega)\, \omega + \epsilon_k}{(Z (\omega)\,
\omega)^2 - \epsilon^2_k - \phi^2(\theta, \omega)},
\label{nine}
\end{equation}
with $\epsilon_k = k^2/2m-\mu$, $Z(\omega)$ the renormalization parameter 
and $\phi\, (\theta, \omega)
= \phi\, (\omega) \cos (2\theta)$ the gap parameter. The Eliashberg
equations for $Z(\omega)$ and $\phi\, (\omega)$ are
\begin{subequations}
\begin{eqnarray}
&(1-Z(\omega))\, \omega = \lambda_z \int^\infty_0 d\omega^\prime \int 
\frac{d\theta}{2\pi}~~~~~\nonumber\\
&{\rm Re}\left\lbrace \left(\frac{Z(\omega^\prime)
\, \omega^\prime}{\left[(Z(\omega^\prime)\, \omega^\prime)^2 - \phi^2
(\omega^\prime) \cos^2 2\theta\right]^{\frac{1}{2}}}\right)\right. \times 
\nonumber\\
& \left.\left(\frac{1}{\omega^\prime + \omega + \Omega_0 - i\, \delta} 
- \frac{1}{\omega^\prime - \omega + \Omega_0 - i\, \delta}\right)\right\rbrace
\label{ten}
\end{eqnarray}
\begin{eqnarray}
& \phi\, (\omega) = \lambda_\phi \int^\infty_0 d\omega^\prime \int
\frac{d\theta}{2\pi}~~~~~~~~\nonumber\\
&{\rm Re}\left\lbrace\left(\frac{\phi\, (\omega^\prime)
\cos^2 2\theta}{\left[(Z(\omega^\prime) \omega^\prime)^2 -
\phi^2(\omega^\prime) \cos^2
2\theta\right]^{\frac{1}{2}}}\right)\right. \times \nonumber\\
& \left.\left(\frac{1}{\omega^\prime + \omega + \Omega_0 - i\, \delta} +
\frac{1}{\omega^\prime - \omega + \Omega_0 - i\, \delta}\right)\right\rbrace,
\label{eleven}
\end{eqnarray}
\end{subequations}
with $\lambda_z = 2|g_q|^2 N(0)/\Omega_0$ and $\lambda_\phi =
2|g_\phi |^2 N(0)/\Omega_0$. Here $N(0)$ is the one-electron density of
states at the Fermi surface.

In order to determine the effect of the phonon on $Z(\omega)$ and
$\phi\, (\omega)$, we will adapt an approximation used in the early studies of
the role of phonons on the superconducting I(V) characteristic.\cite{SA64} 
From the form of eqs \eqref{ten} and
\eqref{eleven}, one sees that there will be structure in $Z(\omega)$ and
$\phi\, (\omega)$ when $\omega\simeq \pm\, (\Omega_0 + \Delta (\theta))$. 
In this case, $\omega^\prime$ will be of order the gap $\Delta(\theta)$ at
the gap edge
\begin{equation}
\Delta(\theta)=\frac{\phi(\theta, \omega = \Delta (\theta))}{Z(\theta,
\omega=\Delta (\theta))}.
\label{twelve}
\end{equation}
Therefore, if the low-energy response in the superconducting state is well
described in terms of BCS $d$-wave quasiparticles, one can replace
$Z(\omega^\prime)$ and $\phi(\theta, \omega^\prime)/Z(\omega^\prime)$
inside the integrals by $Z(0)$ and $\Delta(\theta)=\Delta_0 \cos 2\theta$.
Then, taking the
imaginary parts of eqs \eqref{ten} and \eqref{eleven} we have for
$\omega>0$ 
\begin{subequations}
\begin{eqnarray}
\omega Z_2(\omega)  =  4 \lambda_z \int^{\frac{\pi}{4}}_{\theta_c}
d\theta \frac{(\omega-\Omega_0)}{\left[(\omega-\Omega_0)^2 - \Delta^2_0 \cos^2
2\theta\right]^{\frac{1}{2}}}\label{thirteen}\\
\phi_2(\omega)  =  4 \lambda_\Delta \int^{\frac{\pi}{4}}_{\theta_c}
d\theta \frac{\Delta_0 \cos^2
2\theta}{\left[(\omega - \Omega_0)^2 - \Delta^2_0 \cos^2
2\theta\right]^{\frac{1}{2}}}
\label{fourteen}
\end{eqnarray}
\end{subequations}
Here $\theta_c$ is such that $\Delta (\theta_c) =
\omega-\Omega_0$ and $\phi_2(\omega)$ and $Z_2 (\omega)$ are even
functions of $\omega$ for a time-ordered zero temperature Green's function. 

\begin{figure}[h]
\begin{center}
\epsfig{file=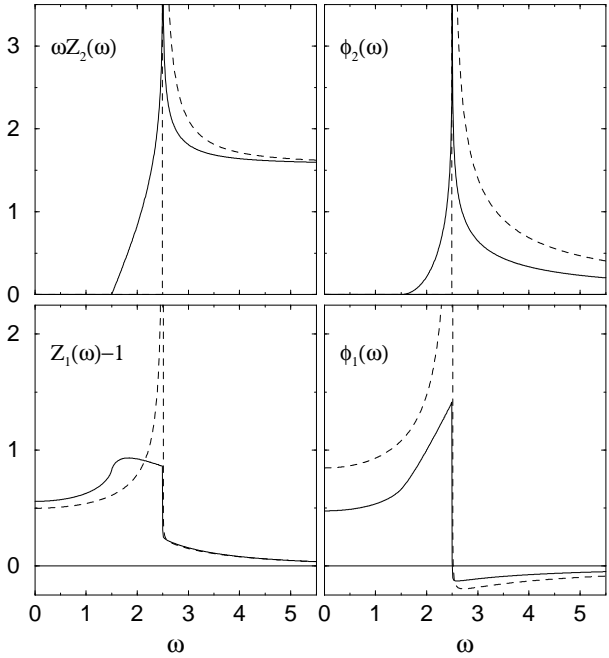,width=8cm}
\caption{Results for the real and imaginary phonon-induced contribution to
$Z(\omega)$ and $\phi(\omega)$ for a $d$-wave (solid) and an $s$-wave
(dashed) superconductor. Here we have taken a cylindrical Fermi surface and
a separable interaction. $Z(\omega)$
and $\phi(\omega)$ are normalized with respect to the appropriate coupling
constant, $\lambda_\phi$ and $\lambda_z$ for the $d$-wave gap and
$\lambda_z$ for
the $s$-wave case. For all of the circular Fermi surface plots, energy is
measured in units of $\Delta_0$ and $\Omega_0=1.5\, \Delta_0$. With this
normalization $\omega Z_2(\omega)$ goes to $\pi/2$ as $\omega \to\infty$.}
\end{center}
\end{figure}

Results for $\omega Z_2(\omega)$ and $\phi_2(\omega)$ are shown in the top 
panel of Fig.~1 for
both a $d_{x^2-y^2}$-wave and an $s$-wave gap with $\Omega_0 = 1.5 \Delta_0$. 
For an
$s$-wave gap, $\cos 2\theta$ is set to 1 and $\theta_c=0$ in eqs
\eqref{thirteen} and \eqref{fourteen}. For the $s$-wave case, the imaginary
parts of $Z(\omega)$ and $\phi(\omega)$ onset when $\omega$ exceeds $\pm
\ (\Omega_0 + \Delta_0)$ and exhibit a square root singularity. For a
$d_{x^2-y^2}$-gap, these functions onset linearly at $\omega=\pm\ \Omega_0$
because of the gap nodes and there is a log singularity at $\pm\ (\Omega_0 +
\Delta_0)$. The real
parts of $Z(\omega)$ and $\phi\, (\omega)$ are obtained from the usual
dispersion relations, and results for $Z_1 (\omega)$ 
and $\phi_1(\omega)$ are shown in the lower panel of Fig.~1.
For the $s$-wave case, $\phi_1$ and $Z_1$ exhibit square
root singularities as $\omega$ approaches $\pm\ (\Omega_0 + \Delta_0)$. This
is just the expected Kramers-Kronig transform of the square root
singularity in $\phi_2$ and $Z_2$. Similarly, the results for 
$\phi_1$ and $Z_1$ for
the $d_{x^2-y^2}$ case exhibit step discontinuities at $\omega=\pm\ (\Omega_0
+ \Delta_0)$ arising from the log singularities in $\phi_2$ and $Z_2$.
Naturally in real materials, phonon dispersion, impurity scattering, and
finite temperature effects broaden these features. Nevertheless, they
provide a simple framework for analyzing the ARPES data. 

\begin{figure}
\begin{center}
\epsfig{file=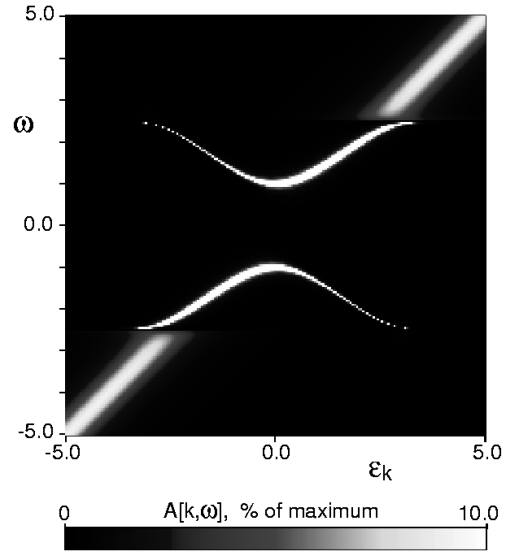,width=6.5cm}
\caption{An intensity plot of the spectral weight $A(k, \omega)$ for the
case of an $s$-wave superconductor coupled to a phonon with frequency
$\Omega_0$. As indicated on the color scale, the figure is `overexposed',
i.e.~features exceeding 10\% of the maximum intensity appear white, in
order to show the weaker features.}
\end{center}
\end{figure}

\begin{figure}
\begin{center}
\epsfig{file=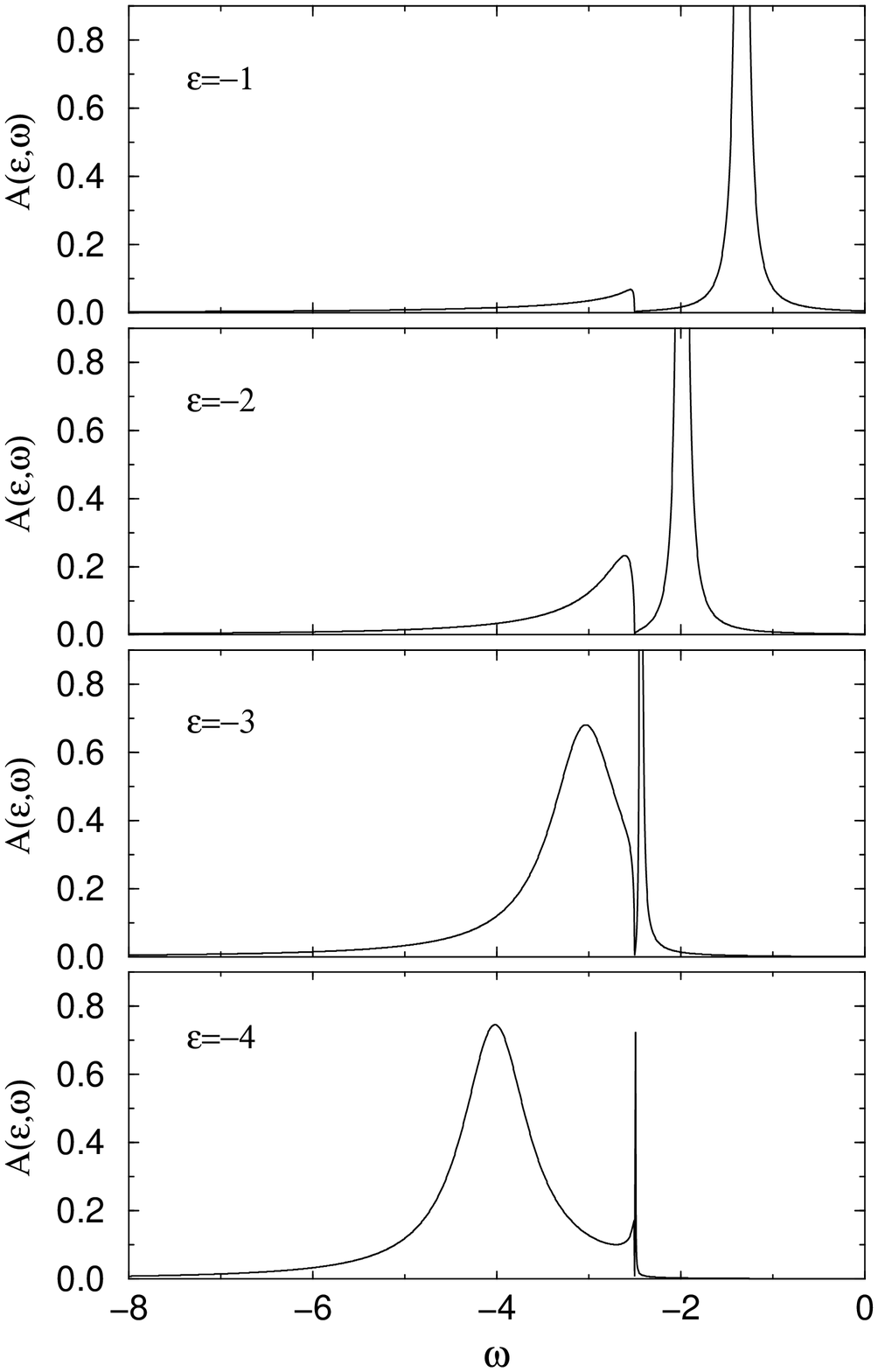,width=7cm}
\caption{Energy distribution curves showing $A(k, \omega)$ versus $\omega$
for various values of $\epsilon_k$ for an $s$-wave superconductor with
$\lambda_z = 0.5$. Here one sees that as $\epsilon_k$ exceeds $\Omega_0
+ \Delta_0 = 2.5$, a peak is left behind whose intensity weakens as
$\epsilon_k$ increases.}
\end{center}
\end{figure}

An intensity plot of $A(k, \omega)$ for the case of an $s$-wave gap is
shown in Fig.~2. Here, $A(k, \omega)$ is obtained from the imaginary part of
$G(k, \omega)$, using the $s$-wave results for $Z(\omega)$ and
$\phi(\omega)$ shown in Fig.~1 with  
$\lambda = 0.5$. The real part of the gap function is
supplemented by an additional contribution from an underlying pairing
interaction so that the magnitude of the gap at the gap edge is equal to
$\Delta_0$.  Results for both the ARPES accessible region
$\omega \leq 0$ and the inverse photoemission region $\omega > 0$ are shown. 
The shift of spectral weight due to the quasiparticle coherence factors
$\frac{1}{2}(1+\frac{\epsilon_k}{E_k})$ is clearly seen as is the
Englesberg-Schrieffer signature showing the asymptotic approach of a 
peak in the spectral function to $\pm
\ (\Omega_0 + \Delta_0)$. Because of the square root singularity in $Z$ and
$\phi$, the asymptotic approach of this peak to $\pm\ (\Omega_0 + \Delta_0)$
varies as $(\lambda\Omega_0/\epsilon_k)^2$. In addition, \cite{note1} 
the Fermi velocity
is renormalized by $Z_1(\Delta_0)\cong 1.3$ so that the dispersion of the peak
for $\omega$ near $\Delta_0$ varies as 
$\sqrt{(\epsilon_k/Z_1(\Delta_0))^2 + \Delta^2_0}$
while for $\omega$ large compared to $\Omega_0$, a
broadened quasiparticle peak disperses as $\epsilon_k$.
Energy distribution curves (EDC) showing $A(k, \omega)$ versus $\omega$ for
various values of $\epsilon_k$ are shown in Fig.~3 for $\omega \leq 0$.
This is the type of EDC that one would expect to see for a traditional
$s$-wave electron-phonon superconductor with a single dominant Einstein
mode.\cite{Sca69} 
More generally, one would have multiple phonon modes and their
dispersion along with possible finite temperature effects would lead to a
richer response.

\begin{figure}
\begin{center}
\epsfig{file=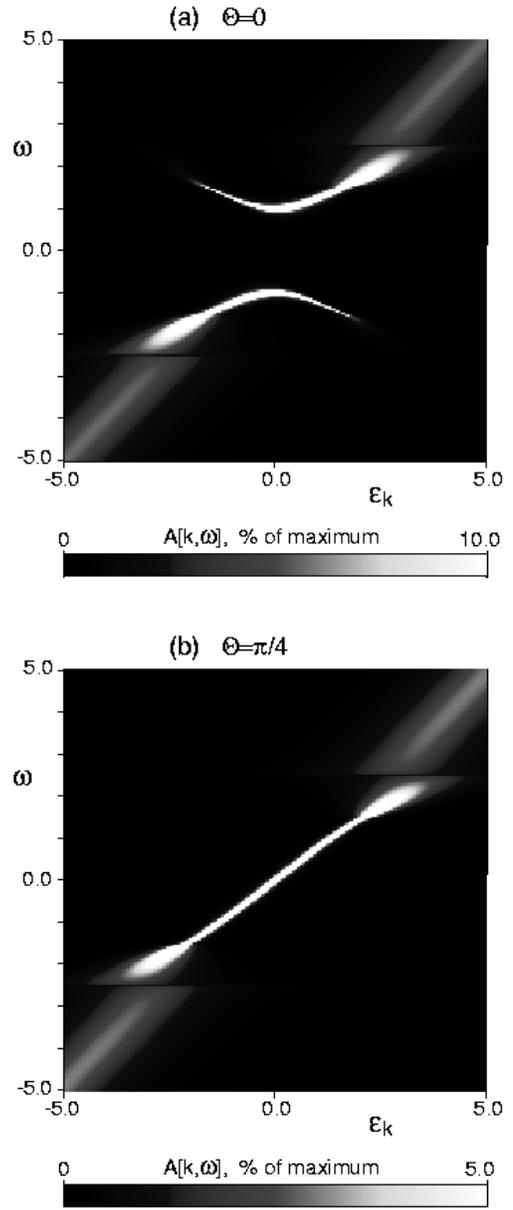,width=6.5cm}
\caption{An intensity plot of $A(k, \omega)$ for a $d$-wave gap. Results
for a $\theta=0$ cut are shown in (a) and in (b) for $\theta=\pi/4$, a
nodal cut.}
\end{center}
\end{figure}

\begin{figure}
\begin{center}
\epsfig{file=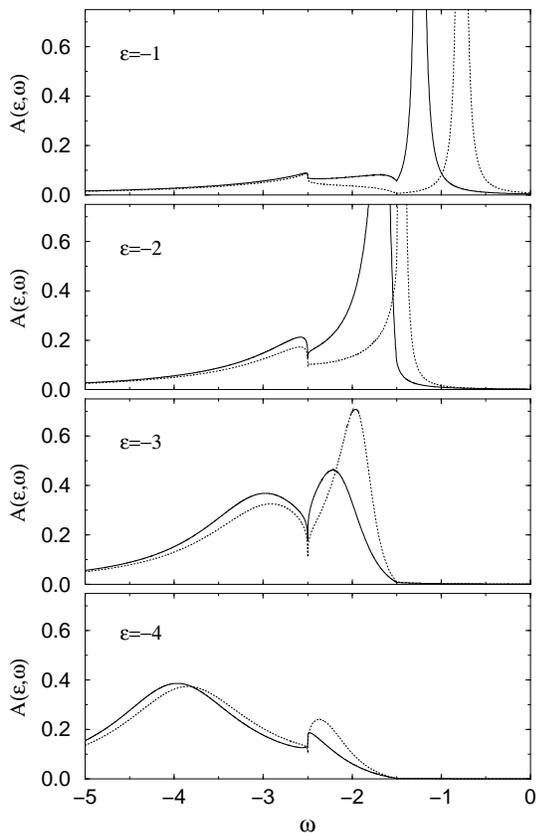,width=7cm}
\caption{Energy distribution curves (EDC) showing $A(k, \omega)$ versus
$\omega$ for different values of $\epsilon_p$ with $\lambda_z =
\lambda_\phi = 0.5$. Results for the $\theta=0$ cut are shown as the solid
curves and for $\theta=\pi/4$ as the dashed curves.}
\end{center}
\end{figure}

Intensity plots of $A(k, \omega)$ for the case of a $d_{x^2-y^2}$ gap are
shown in Fig.~4. Just as for the $s$-wave case, $\phi_1(\theta, \omega)$ is
supplemented so that the gap at the gap edge is $\Delta_0 \cos 2\theta$.
Fig.~4(a) shows $A(k, \omega)$ for a cut along the
antinodal direction in $k$-space $(\theta=0)$, while Fig.~4(b) shows the
results for a cut along the nodal direction $(\theta = \pi/4)$. The
antinodal cut resembles the $s$-wave case in the transfer of spectral
weight as $\epsilon_k$ passes through the Fermi energy and the
renormalization of the quasiparticle dispersion. However, the
Englesberg-Schrieffer signature no longer asymptotically approaches $\pm
\ (\Omega_0 + \Delta_0)$, but rather appears to be broadened and cut off.
In the $s$-wave case, the broadening
due to the electron-phonon interaction did not set in until $|\omega|$
exceeded $\Omega_0 + \Delta_0$ leading to the long sweep of the
peak which occurs for $|\omega|$ just below $(\Omega_0 + \Delta_0)$.
However, the nodal regions associated with a $d_{x^2-y^2}$ gap lead to a
finite broadening when $|\omega|$ exceeds $\Omega_0$. The onset of this
broadening is seen by the faint horizontal line in Fig.~4 where the
intensity changes from black to blue at larger values of $\epsilon_k$. 
As we will discuss, termination of this peak is a reflection of the fact
that for a $d_{x^2-y^2}$-gap, $Z_1$ and $\phi_1$ have step discontinuities
at $\pm\ (\Omega_0 + \Delta_0)$ rather than the square root singularities
associated with an $s$-wave gap.  

The nodal cut, shown in Fig.~4(b), appears on first glance to be similar to
what one would expect for the normal state.  That is, a renormalized
$\epsilon_k/Z_1(k, 0)$ dispersion for $\omega \ll \Omega_0$ with the
dispersion returning to its band value $\epsilon_k$ for $\omega \gg
\Omega_0$. However, the cut-off Englesberg-Schrieffer signature still occurs 
for $|\omega|=\Omega_0 + \Delta_0$. Thus, the full antinodal gap $\Delta_0$ 
enters as the
characteristic kink energy for all momentum slices. This simply reflects
the $|\omega|=\Omega_0+\Delta_0$ singularities in $Z$ and $\phi$ shown in
Fig.~1. Again, the broadening of the Englesberg-Schrieffer peak when
$|\omega|$ exceeds $\Omega_0$ is clearly seen in Fig.~4(b).
In Fig.~5, various EDC slices of $A(k, \omega)$ are shown for the
$d_{x^2-y^2}$ case. Comparing these with the $s$-wave case, one sees the
broadening and truncation of the Englesberg-Schrieffer lower peak.

\begin{figure}
\epsfig{file=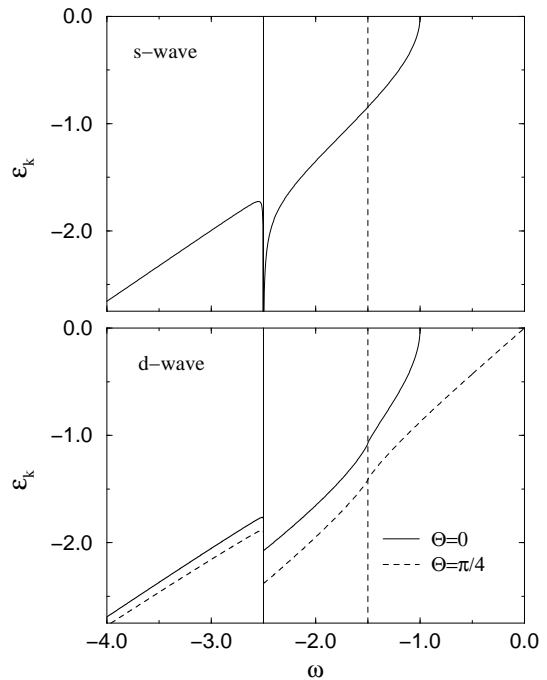,width=7cm}
\caption{Plots of $\epsilon_k = -\sqrt{(Z_1 (\omega) \omega)^2 - \phi^2_1
(\omega, \theta)}$ versus $\omega$ along the negative $\omega$ axis for
$\lambda_\alpha = 0.5$ show the structure of the Englesberg-Schrieffer
signature. The upper panel is for an $s$-wave gap and the lower panel is
for a $d$-wave gap with $\theta=0$ shown as the solid curve and
$\theta=\pi/4$ as the dashed curve.}
\end{figure}

The difference in the structure of the Englesberg-Schrieffer signature
between the $s$- and the $d_{x^2-y^2}$-cases can be understood from the
plots of
\begin{equation}
\epsilon_k = - \sqrt{(Z_1(\omega) \omega)^2 - \phi^2_1 (\omega)}
\label{fifteen}
\end{equation}
shown in Fig.~6. One can see that as one probes
$\epsilon_k$ states which are further below the Fermi energy, two solutions
of Eq.~\eqref{fifteen} develops. For the $s$-wave case shown in the upper
panel of Fig.~6, an undamped lower
energy branch asymptotically approaches $\omega = - (\Omega_0 +
\Delta_0)$, and a second quasiparticle branch at $\omega\simeq -
\epsilon_k$ evolves which is damped by the imaginary parts of $Z$ and
$\phi$. As we have seen, these branches are reflected in the structure of
$A(k, \omega)$ and the lower energy branch represents the characteristic
Englesberg-Schrieffer signature for an $s$-wave superconductor. Similar
plots for the $d_{x^2-y^2}$-case with
$\theta=0$ and $\theta = \pi/4$ are shown in the lower panel of Fig.~6.
Here, unlike the $s$-wave case, the low energy branch is
terminated, reflecting the fact that the singularities in $Z_1$ and $\phi_1$ for
the $d$-wave case are simply step discontinuities at $\pm\ (\Omega_0 +
\Delta_0)$. The onset of damping processes for the $d_{x^2-y^2}$ case when
$\omega < -\Omega_0$ give rise to the discontinuity in slope seen at
$\omega= -\Omega_0$.

\section{Band Structure and the Effect of a Momentum-Dependent Coupling}

We turn next to the effects of the band structure and to the 
momentum dependence of the electron-phonon coupling. For the band
structure,
consider a square lattice with a near-neighbor hopping $t$ and a
next-near-neighbor hopping $t^\prime$. In this case
\begin{equation}
\epsilon_k = - 2t (\cos k_x + \cos k_y) - 4t^\prime \cos k_x \cos k_y -
\mu.
\label{sixteen}
\end{equation}
For $t^\prime/t = -0.3$ and $\mu/t = -1$, one has the typical Fermi surface
shown in Fig.~7 and the single spin electron density of states shown in
the inset. We take the gap to be
\begin{equation}
\Delta_k = \Delta_0 (\cos k_x - \cos k_y)/2.
\label{seventeen}
\end{equation}
In this case, the one-electron Green's function can be written in the form
\begin{equation}
G(k, \omega) = \frac{Z(k, \omega)\omega + (\epsilon_k + X(k,
\omega))}{(Z(k, \omega)\omega)^2 - (\epsilon_k + X(k, \omega))^2 - \phi^2
(k, \omega)}.
\label{eighteen}
\end{equation} 
Adopting the same approximation as before, the phonon-induced
contributions to the imaginary parts of the renormalization, energy shift,
and gap parameters are given by
\begin{subequations}
\begin{eqnarray}
\omega Z_2 (k, \omega) & = &  \frac{\pi}{2N}
\sum\limits_{k^\prime} |g(k-k^\prime)|^2\times~~~~~\nonumber\\
\Bigl(\delta(E_{k^\prime}& + & \Omega_0 - \omega) -   \delta
(E_{k^\prime} + \Omega_0 + \omega)\Bigr),
\label{nineteen}
\end{eqnarray}
\begin{eqnarray}
X_2 (k, \omega)  = - \frac{\pi}{N}\ \sum\limits_{k^\prime}
|g(k-k^\prime)|^2\ \frac{\epsilon_{k^\prime}}{2E_{k^\prime}}\times~~~~~
\nonumber\\
\Bigl(\delta\left(E_{k^\prime} + \Omega_0 - \omega\right) + \delta
\left(E_{k^\prime} + \Omega_0 + \omega\right)\Bigr),
\label{twenty}
\end{eqnarray}
\begin{eqnarray}
\phi_2 (k, \omega)  =  \frac{\pi}{N}\ \sum\limits_{k^\prime}
|g(k-k^\prime)|^2\ \frac{\Delta_{k^\prime}}{2E_{k^\prime}}\times~~~~~
\nonumber\\
\Bigl(\delta\left(E_{k^\prime} + \Omega_0 - \omega\right) + \delta
\left(E_{k^\prime} + \Omega_0 + \omega\right)\Bigr).
\label{twentyone}
\end{eqnarray}
\end{subequations}
Here, as before, we assume that an underlying pairing interaction, most
likely spin-fluctuations, gives rise to a zero temperature $d_{x^2-y^2}$
superconducting state. At low energies this state is characterized by a
renormalized band structure Eq.~\eqref{sixteen} and chemical potential, a
renormalized coupling constant $g(q)$, and a gap given by
Eq.~\eqref{seventeen}. These parameters have been used in the Eliashberg
equations to describe the state which enters when an excitation at energy
$\omega >\Omega_0$ decays to a lower energy $E_{k^\prime}$ state (or when 
$\omega < - \Omega_0$ decays to $-E_{k^\prime}$). The real parts of $Z(k,
\omega)$, $\phi(k, \omega)$, and $X(k, \omega)$ are again found from
the Kramer-Kronig dispersion relation. The spectral weight $A(k, \omega)$
is then obtained from Eq.~\eqref{one} with the chemical potential shift
removed from $X_1(k, \omega)$ and a contribution added to the real part of
the gap so that the real part of the gap at the gap edge remains equal 
to $\Delta_k$, Eq.~\eqref{seventeen}. Note that the contributions of the
underlying pairing interaction to $Z$ and $X$, as well as the higher energy
part of $\phi$, have not been included. Thus, there are additional
renormalization and damping effects which do not appear. We basically are 
seeking to understand the leading contribution of
the electron-phonon interaction which is superimposed on top of the other
many-body interactions. This approach rests on the idea that in the
superconducting state the low-lying electronic states are well described by
BCS $d_{x^2-y^2}$ excitations \cite{Hof02}
with renormalized band parameters $t$,
$t^\prime$, and $\mu$, a $d_{x^2-y^2}$-wave gap $\Delta_k$, and
renormalized electron-phonon coupling constants.  Note, that here we are
not taking into account the possible change in $q$-dependence of the
electron-phonon couplings produced for example by the Hubbard 
$U$.\cite{ZK96,Hua03} 

\begin{figure}
\begin{center}
\epsfig{file=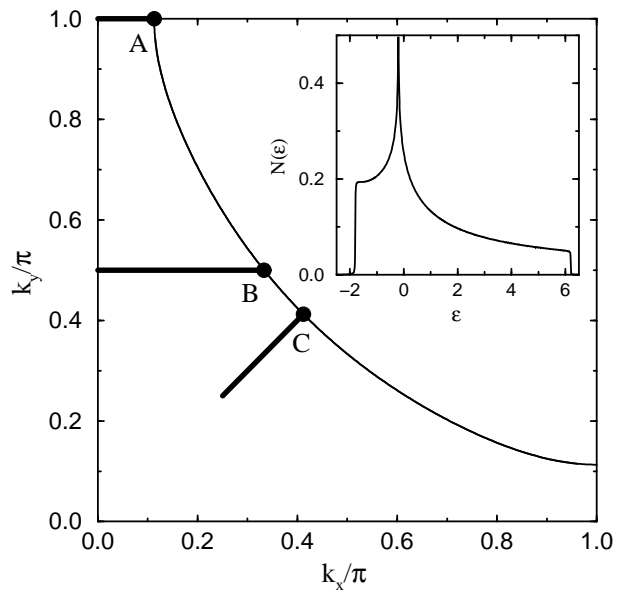,width=8cm}
\caption{The Fermi surface for $t^\prime/t=-0.3$ and $\mu=-1$. The inset
shows the single spin electron density of states.  We will discuss the
spectral weight for the cuts marked $A$, $B$, and $C$.}
\end{center}
\end{figure}

\begin{figure}
\begin{center}
\epsfig{file=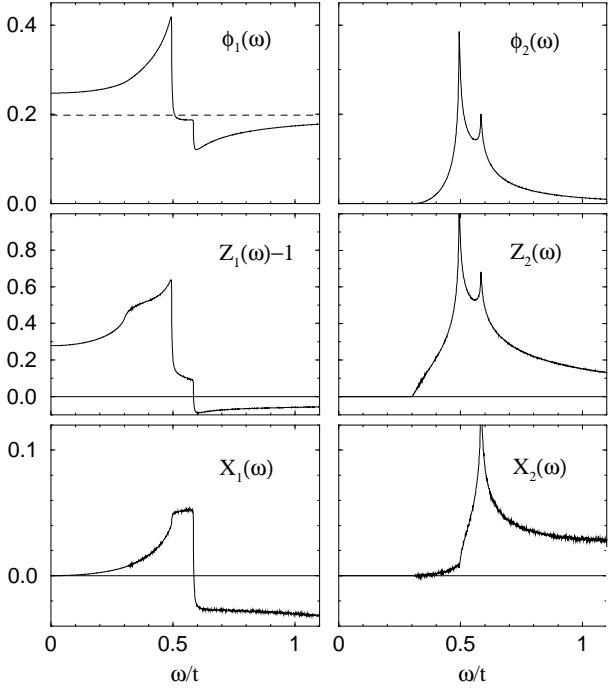,width=8cm}
\caption{The self-energy parameters $\phi$, $Z$, and $X$ versus $\omega$
for the case of the buckling phonon coupling, Eq.~\eqref{three}, with ${\bf k}
= {\bf k}_A$ corresponding to the point $A$ of Fig.~7. Here $|g|^2=0.5$.}
\end{center}
\end{figure}

We begin by looking at the self-energy terms for the case of the buckling
mode with $|g(q)|^2$ given by Eq.~\eqref{three} and $|g|^2 = 0.5$ in units
of $t^{-2}$. Results for $\phi(\omega, k)$, $Z(\omega, k)$, and $X(\omega,
k)$ are shown in Fig.~8 for $k$ at point $A$ shown in Fig.~7. 
The imaginary parts of $Z$ and $\phi$ exhibit the expected log singularity at
$\Delta_0 + \Omega_0$ that we previously saw for the case of a circular
Fermi surface. In addition, there is a second log singularity at 
$E(0, \pi) + \Omega_0$ with $E(0, \pi) = \sqrt{\epsilon^2 (0, \pi) + 
\Delta^2_0}$ which comes from the Van
Hove singularity\cite{EN00} at $k =(0, \pi)$. These log singularities in
$Z_2$ and $\phi_2$ manifest themselves via the Kramers-Kronig dispersion 
relation as
step-down discontinuities in $Z_1$ and $\phi_1$, as seen in Fig.~8. 
The energy shift parameter $X$ has only the Van Hove singularity. Naturally, 
the dispersion of the phonon mode as well as finite temperature and lifetime
effects will broaden these features in the actual system.
The energy distribution of the spectral weight $A(k, \omega)$ for the
buckling mode at momentum $k_A$ is plotted in Fig.~9. It shows the
quasiparticle peak at the gap edge $\Delta_{k_A}$ as well as structure
associated with the buckling phonon at $\Omega_0 + \Delta_0$ and
$\Omega_0 + E(0, \pi)$.  

As noted in the introduction, one would like to determine whether the
structure observed in the ARPES data is due to phonons or the
$\pi$-resonance spin fluctuation mode. Eschrig and Norman\cite{EN00,EN03}
have analyzed the effect of the $\pi$-resonance using a detailed tight
binding fit of the band energy $\epsilon_k$ and a coupling to the
$\pi$-resonant mode of frequency $\Omega_0$ given by
\begin{equation}
|g(q)|^2 = g^2_{SF}\ \frac{w_Q}{1+ 4\xi^2 [\cos^2(q_x/2) + 
\cos^2(q_y/2)]}.
\label{twentytwo}
\end{equation}
Here, we will use the $t-t^\prime$ band structure of Eq.~\eqref{sixteen}
with $t^\prime/t = -0.3$ and $\mu=-1$,
set $w_Q=1$, $\xi=2$, and set $g^2_{\rm SF} = 5$ which corresponds to
having $\frac{3}{4}(\frac{\bar U}{t})^2 = 5$ in an effective Hubbard RPA
interaction. In addition, with this choice for $g^2_{\rm SF}$ we will find
that $Z_1(k_F, 0)$ at the nodal point $C$ is comparable with $Z_1(k_F, 0)$ for
the phonons. This makes it convenient for addressing the question of
whether there are significant spectral differences due simply to the
structure of the momentum-dependent couplings that would allow one to
determine the nature of the mode from the ARPES data. \cite{note2}
Note that for the spin-fluctuation interaction with $|g(q)|^2$ given by 
Eq.~(\ref{twentytwo}), there is a minus sign on the righthand side of 
Eq.~(\ref{twentyone}) for the gap parameter. For the three types of phonon 
couplings we take $|g|^2=0.5$ in units of $t^{-2}$. This gives $Z_1(k_F, 0) 
\simeq 1.3$ corresponding to an effective \cite{ASJL96,JADS98} $\lambda 
\sim 0.3$. For the $\pi$-resonance mode coupling, setting $g_{SF}^2=5$ 
gives $Z_1(k_F, 0) \simeq 1.3$. 

\begin{figure}
\epsfig{file=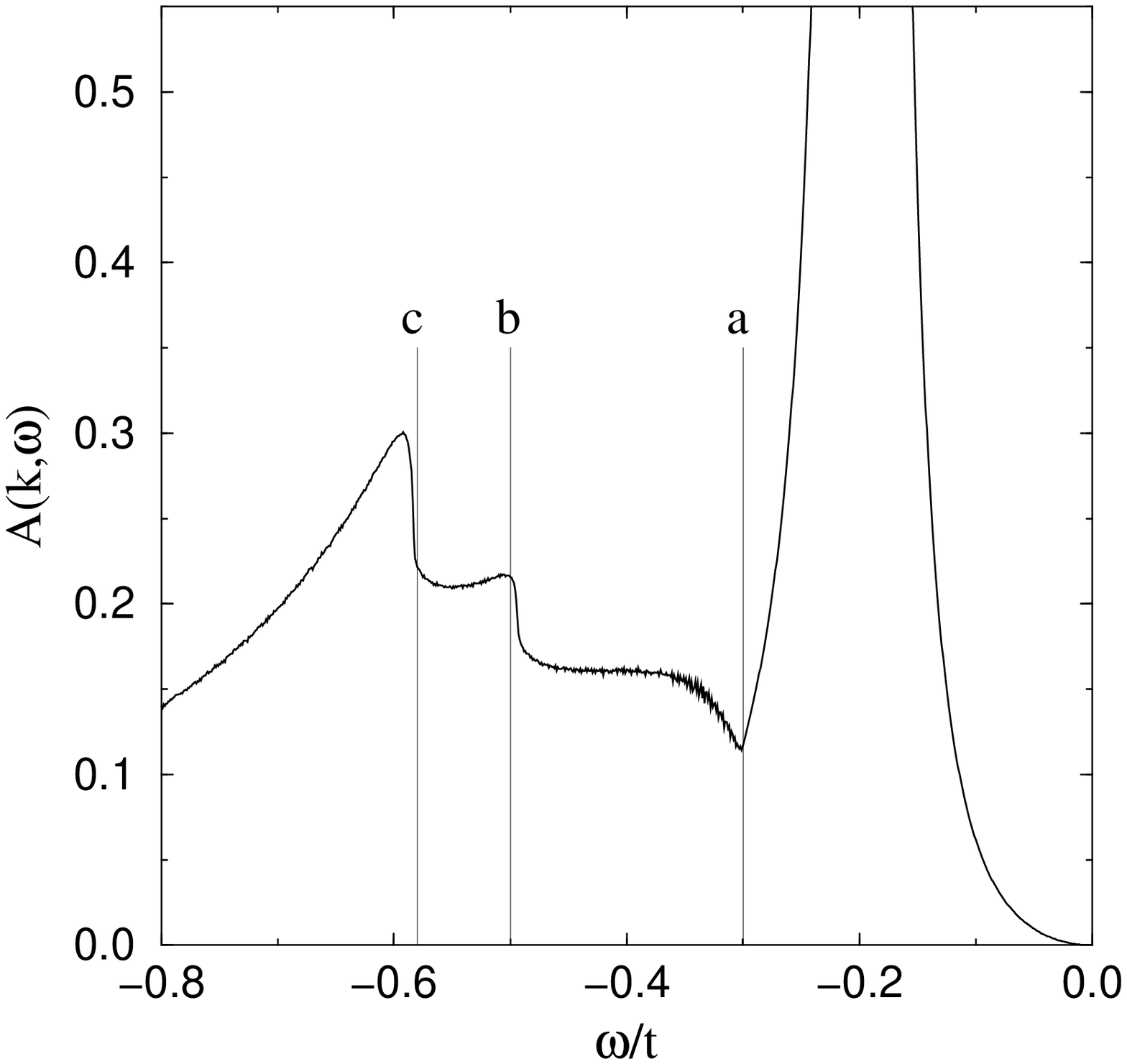,width=8cm}
\caption{$A(k, \omega)$ versus $\omega$ at ${\bf k} = {\bf k}_A$ for the
case of a buckling mode with $|g|^2 = 0.5$. With energy measured 
in units of $t$,
the gap amplitude $\Delta_0=0.2$, and the phonon energy $\Omega_0=0.3$. The
vertical lines $a$, $b$, and $c$ mark the characteristic energies 
$-\Omega_0$, $-(\Omega_0
+ \Delta_0)$, and $-(\Omega_0 + E(0,\pi))$ respectively.}
\end{figure}

\begin{figure}
\begin{center}
\epsfig{file=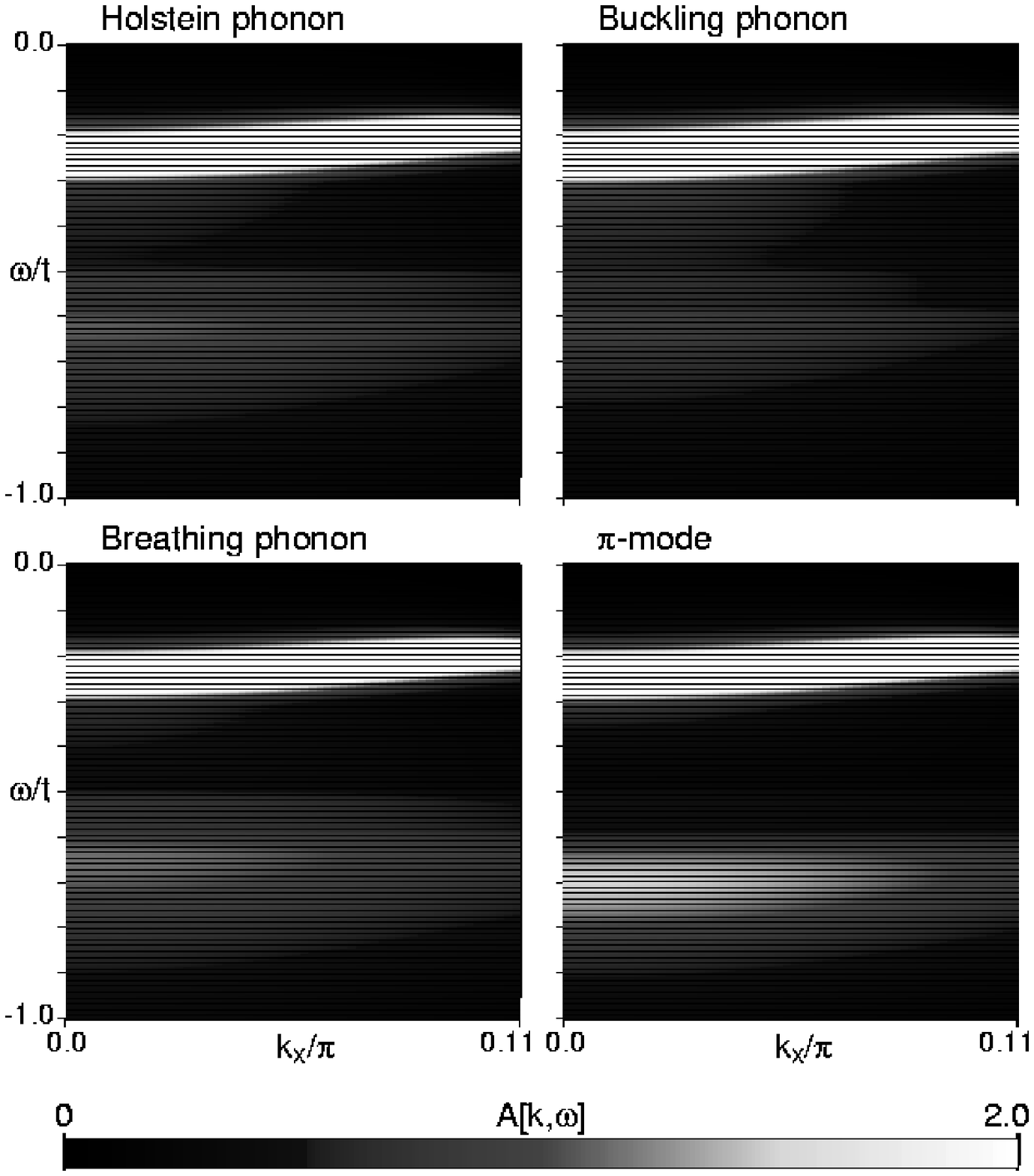,width=8cm}
\caption{
Intensity plots of $A(k, \omega)$ for the momentum cut $A$ for
four different couplings corresponding to the Holstein mode with $|g(q)|^2$
constant, the buckling mode Eq.~\eqref{three}, breathing mode
Eq.~\eqref{five}, and $\pi$-resonance magnetic mode coupling
Eq.~\eqref{twentytwo}. Here, $|g|^2=0.5$ and $g^2_{SF}=5$. The
cut-off indicated on the color scale refers to the actual spectral weight
intensity (as opposed to the relative scale used in the previous intensity
plots) so that one can directly compare the effects of the different 
couplings.}
\end{center}
\end{figure}

\begin{figure}
\begin{center}
\epsfig{file=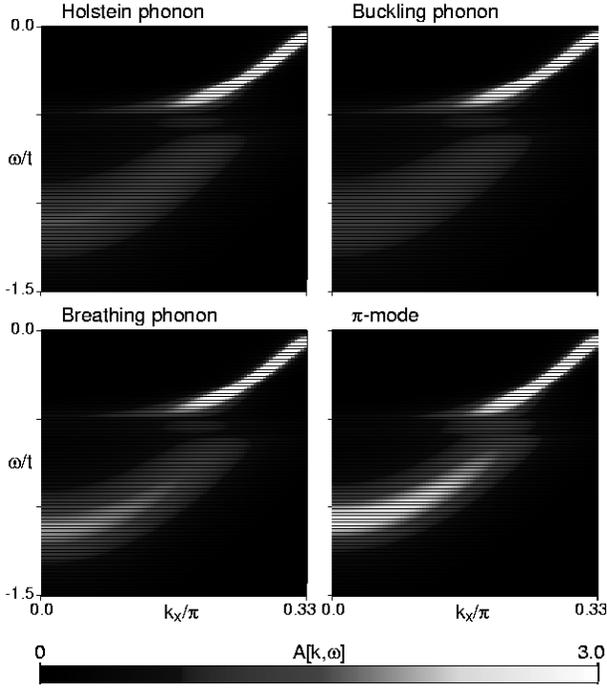,width=8cm}
\caption{Intensity plots of $A(k, \omega)$ for the momentum cut $B$ for the
four different couplings.}
\end{center}
\end{figure}

Intensity plots of $A(k, \omega)$ for the constant Holstein coupling, the
buckling mode coupling Eq.~\eqref{three}, the breathing mode coupling
Eq.~\eqref{five}, and the $\pi$-resonance mode coupling
Eq.~\eqref{twentytwo}, are shown in Fig.~10 for the momentum cut $A$.
Similar intensity plots for the momentum cuts $B$ and $C$ are shown in Figs
11 and 12. In Fig.~10, one sees a high intensity quasiparticle peak and
weaker structures onsetting at $\omega=-(\Omega_0 + \Delta_0)$ and
$-(\Omega_0 + E(0, \pi)$ due to the coupling to the phonon or magnetic
resonance modes. For the $B$ momentum cut shown 
in Fig.~11, one can now move deep
enough inside the Fermi sea that the Englesberg-Schrieffer lower energy
peak (the upper bright curve in the figures) is broadened when $\omega$
becomes less than $-\Omega_0$ and terminated at a finite value of $k_x$ as
$\omega$ approaches $-(\Omega_0 + \Delta_0)$. At still higher energies
($\omega$ more negative), a damped
quasiparticle branch is seen. The nodal $C$ cut is shown in Fig.~12. Here,
one clearly sees the Englesberg-Schrieffer signature with a 
quasiparticle peak which varies as $\epsilon_k/Z_1 (k_F, 0)$ near the Fermi
surface, then disperses and bends as $\omega$ approaches $-(\Omega_0 +
\Delta_0)$. This peak is then terminated as a broadened high energy
quasiparticle branch appears at more negative values of $\omega$.   

\begin{figure}
\begin{center}
\epsfig{file=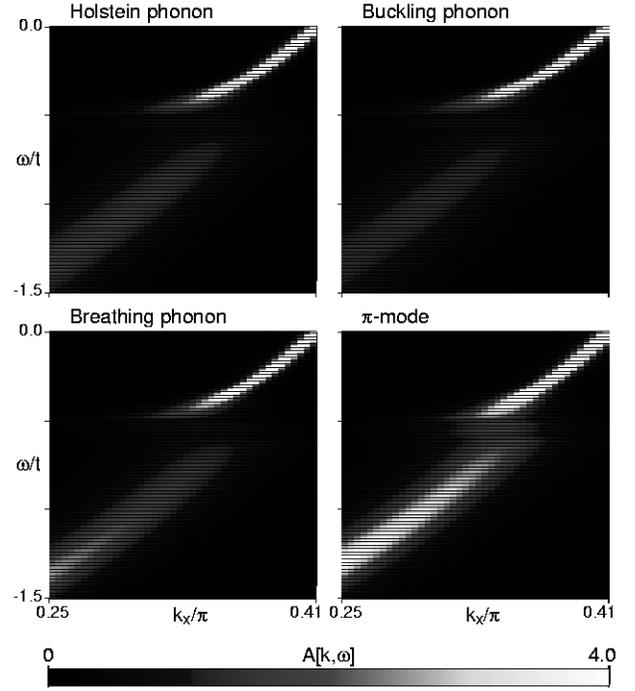,width=8cm}
\caption{Intensity plots of $A(k, \omega)$ for the nodal momentum cut $C$
for the four different couplings.}
\end{center}
\end{figure}

\begin{figure}
\begin{center}
\epsfig{file=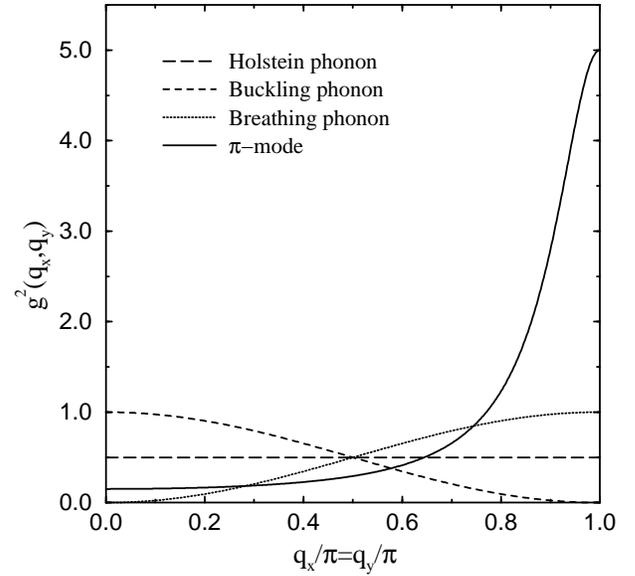,width=8cm}
\caption{The momentum dependence of the $|g(q)|^2$ coupling versus
$q=q_x=q_y$ for the four different modes with $|g|^2=0.5$ for the phonon
modes and $g^2_{SF}=5$ for the $\pi$-mode. Note that this is a slice of a
two-dimensional $(q_x, q_y)$ surface and that the volume enclosed by these
surfaces is $(2\pi)^2|g|^2$ for each of these couplings.}
\end{center}
\end{figure}

The difference of $A(k, \omega)$ for the various modes is in fact subtle
since all four have an Einstein spectrum with $\Omega_0 = 0.3t$, a
$d_{x^2-y^2}$ gap with $\Delta_0 = 0.2t$, and a band structure with
$t^\prime/t = - 0.3$ and $\mu=-1$. Thus, the characteristic energies
$\Delta_0$, $\Omega_0 + \Delta_0$, and $\Omega_0 + E(0, \pi)$ are the same.
In addition as discussed, we have chosen the coupling constants so that
$|g(q)|^2$ averaged over the Brilloin zone is the same for all four cases.
Thus, the basic difference is the momentum structure of the different
couplings shown in Fig.~13 for $q_x=q_y$. Here, we see that the
spin-fluctuation resonant mode is clearly most strongly peaked at large
momentum, followed by the breathing mode phonon, the uniform Holstein
coupling, and lastly the buckling mode phonon which has $|g(\pi, \pi)|^2=0$.
One consequence of the strong peak in the magnetic reasonance-mode coupling 
is seen in Fig.~10 for the $A$ cut. Here, the increase of the intensity of 
the spectral weight $A(k, \omega)$ which occurs when $\omega$ decreases below 
$-(\Omega_0 + E(\pi, 0))$ is greatest for the spin-fluctuation 
$\pi$-resonance and smallest for the buckling mode.

\begin{figure}
\begin{center}
\epsfig{file=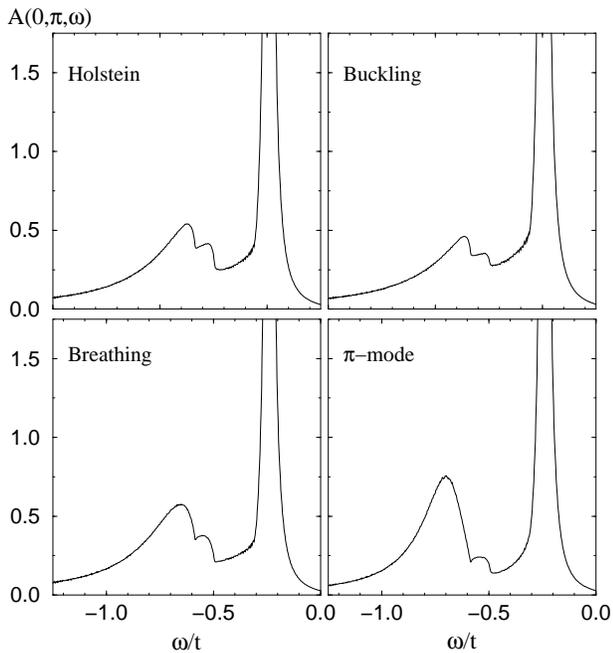,width=8cm}
\caption{Energy distribution curves (EDC) showing $A(k, \omega)$ versus 
$\omega$ at $k=(0, \pi)$ for the four different modes.}
\end{center}
\end{figure}

In Fig.~14 we show the energy distribution curves for the four modes for
momentum 
$k=(0, \pi)$. $A(k, \omega)$ for all of the modes shows a strong peak at
$\Delta_0$. For the $\pi$-mode, this is followed by a dip and then a
secondary peak which develops as $\omega$ decreases below the Van Hove
threshold at $-(\Omega_0 + E(0, \pi))$. It is this peak-dip-hump structure,
for the case in which the effects of the bilayer splitting can be
eliminated, that has been identified as a `fingerprint' of the
$\pi$-resonance.\cite{EN00,EN03,Bor02,Kim03} Here, we see that indeed this 
structure is most pronounced for the $\pi$-mode and smallest for the buckling 
mode. However, this is a quantitative effect rather than a qualitative one 
and if the phonon coupling increases at large momentum transfers, such as 
in the case of the breathing mode, this feature returns although not as 
strongly as for the $\pi$-mode.

\section{Conclusions}

The Englesberg-Schrieffer-like structure in the ARPES data of BISCO is
consistent with the existence of an Einstein-like mode with $\Omega_0 \sim
40$meV coupled to the electrons as suggested by various authors.
\cite{SS97,Nor97,ND98}   However,
it seems that it will be difficult to determine the origin of the mode
based solely upon the $(q_x, q_y)$ momentum dependence of its coupling. 
One might
have thought that the $q$-dependence of the coupling or in the case of the
$\pi$ mode, the $q$-dependence of the resonance that has been parameterized
as a $q$-dependent coupling, would give rise to clearly
identifiable structure in $A(k, \omega)$. However, all of the modes show
quite similar characteristic features at energies $\Omega_0 + \Delta_0$ and
$\Omega_0 + E(0, \pi)$, which appear throughout the zone.

It would appear that the best place to look for a feature that could
distinguish between, for example, the buckling phonon mode and the
$\pi$-resonant mode is near the $k=(0, \pi)$ point. Here, the strong
coupling of the $\pi$ mode to the electrons for $q$ near $(\pi, \pi)$ leads to a
secondary peak onsetting at an energy $\omega= -(E(0, \pi) + \Omega_0)$. For
the buckling phonon mode, the coupling at $q=(\pi, \pi)$ vanishes and there
is only a relatively weak response in this same frequency range. However,
as we have seen, there is a secondary peak for the breathing mode which
has nearly the same strength as that for the $\pi$-mode. Thus, the observed
peak-dip-hump structure could also be consistent with a coupling to the
oxygen-breathing mode. Recently, it has been suggested that the
$q_z$ dependence for a bilayer system may identify the mode as having
$q_z=\pi$, which would provide support for the $\pi$ resonance.\cite{EN02} 
However,
further work on the odd and even bilayer phonon coupling is needed for
comparison.

While the coupling to the $\pi$-resonance mode along with a higher energy
continuum spin-fluctuation spectrum provides an attractive unified
framework, our results leave open the possibility that an oxygen phonon
mode could also play a role. As we
have seen, even with a relatively modest coupling constant $\lambda \sim
0.3$, one would expect to see evidence of some oxygen phonon modes. 
If they are not
seen, then this suggests that the strong Coulomb many-body effects act to
suppress the electron-phonon coupling. Alternatively, if it can be shown
that the $\pi$-mode is not viable, oxygen phonon modes could
provide a source for the resonant mode features. The continuum spin
fluctuations would, of course, also contribute in this mixed scenerio.
Here we should note that even if the mode were identified
as the buckling mode, we find that its contribution to the magnitude of the
$d_{x^2-y^2}$ gap is negligible because the increase in $Z_1$ more than
offsets the increase in $\phi_1$ (in Eq.~\eqref{four} the $d$-wave
coupling term is only 1/4 of the uniform coupling). This is in agreement
with the results of Eliashberg-like $T_c$ calculations \cite{NSB99}
which find that, while the buckling
phonons can provide an attraction in the $d_{x^2-y^2}$-channel, its
contribution to $Z$ leads to an overall reduction in $T_c$.

To conclude, from the results that we have found, it seems likely that the
identification of the excitation responsible for the structure in the ARPES
data will be decided on grounds other than the momentum dependence of the
effective coupling. One aspect that remains under discussion is the
strength of the various couplings. For the O phonon modes, LDA calculations
\cite{ASJL96, JADS98} find intermediate coupling strengths with $\lambda
\sim 0.3$ to 0.5. From our calculations it would appear that at this
strength, one should in fact see structure in $A(k, \omega)$. If this is
not seen, it raises the question of why is the electron-phonon coupling
weakened in strongly-correlated materials? \cite{ZK96,Hua03}
The coupling to the $\pi$-resonance mode would appear
to raise the opposite problem. That is, if the $\pi$-resonance mode is
responsible for the Englesberg-Schrieffer-like signature in the ARPES
spectrum, how can it be coupled so strongly? \cite{EN03,KKA02,ACENS02} 

\acknowledgments

We would like to thank Z.-X.~Shen for discussing his data with us, for his
physical insights, and his enthusiasm for this project. DJS
would also like to also acknowledge very useful discussions 
with S.V.~Borisenko, W.~Hanke, and S.A.~Kivelson. 
AWS would like to acknowledge support from the Academy of Finland under
project No.~26175. DJS acknowledges support from the National Science
Foundation under grant No.~DMR02-11166.


\begin{thebibliography}{99}

\bibitem{Mar90} S.~Martin {\it et al.}, Phys.~Rev.~B {\bf 41}, 846
(1990).

\bibitem{Fra94} J.P.~Franck, in {\sl Physical Properties of High $T_c$
Superconductors IV}, ed.~by D.M.~Ginsberg, World Scientific, Singpore,
p.~189 (1994).

\bibitem{Cra90} M.K.~Crawford, {\it et al.}, Phys.~Rev.~B {\bf 41},
282 (1990). 

\bibitem{BHI00} C.~Boulesteix, K.~Hewitt, J.C.~Irwin,
J.~Phys.~Cond.~Mat. {\bf 12}, 9637 (2000).

\bibitem{MRS92} R.M.~Macfarlane, H.~Rosen, and H.~Seki, 
Sol.~St.~Comm. {\bf 83}, 343 (1992).

\bibitem{Mcq01} R.J.~McQueeney {\it et al.}, Phys.~Rev.~Lett. {\bf 82},
628 (1999).

\bibitem{Pyk93} N.~Pyka {\it et al.}, Phys.~Rev.~Lett. {\bf 70} 
1457 (1993).
 
\bibitem{Lan01} A.~Lanzara {\it et al.}, Nature {\bf 412}, 
510 (2001).

\bibitem{SLIN02} Z.-X.~Shen, A.~Lanzara, S.~Ishihara, and N.~Nagaosa,
Phil.~Mag.~B {\bf 82}, 1349 (2002).

\bibitem{CEKO02} E.W.~Carlson, V.J.~Emery, S.A.~Kivelson, and D.~Orgad,
{\sl Concepts in High Temperature Superconductivity}, cond-mat/0206217.

\bibitem{ES63} S.~Englesberg and J.R.~Schrieffer, Phys.~Rev. {\bf
131}, 993 (1963).

\bibitem{DHS03} A.~Damascelli, Z.~Hussain, and Z.-X.~Shen, 
Rev.~Mod.~Phys. {\bf 75}, 473 (2003).

\bibitem{Kam01} A.~Kaminski {\it et al.}, Phys.~Rev.~Lett. {\bf 86},
1070 (2001).

\bibitem{EN00} M.~Eschrig and M.R.~Norman, Phys.~Rev.~Lett. {\bf 85},
3261 (2000).

\bibitem{EN03}  M.~Eschrig and M.R.~Norman, Phys.~Rev.~B {\bf 67},
144503 (2003). 

\bibitem{ZK96} R.~Zeyher and M.L.~Kulic, Phys.~Rev.~B {\bf 53} 2850
(1996).

\bibitem{SA95} J.~Song and J.F.~Annett, Phys.~Rev.~B {\bf 51}, 3840
(1995); {\it ibid.} {\bf 52}, 6930 (E) (1995).

\bibitem{Sca95} D.J.~Scalapino, J.~Phys.~Chem.~Solids {\bf 56}, 
1669 (1995).

\bibitem{ND96} A.~Nazarenko and E.~Dagotto, {\sl Phys.~Rev.~B} {\bf 53},
R2987 (1996).

\bibitem{DMFT96} T.~Dahm, D.~Manske, D.~Fay, and L.~Tewordt,
Phys.~Rev.~B {\bf 54}, 12006 (1996).

\bibitem{BS96} N.~Bulut and D.J.~Scalapino, Phys.~Rev.~B {\bf 54}, 
14971 (1996).

\bibitem{Hua03} Z.B.~Huang, W.~Hanke, E.~Arrigoni, and D.J.~Scalapino; 
cond-mat/0306131.  These Monte Carlo Calculations for a Hubbard model find
that, while the electron-phonon vertex is suppressed at large momentum
transfers, it can actually be enhanced at small momentum transfers. This
latter effect differs from Ref.~\onlinecite{ZK96}.

\bibitem{KKA02} H.-Y.~Kee, S.A.~Kivelson, and G.~Aeppli,
Phys.~Rev.~Lett. {\bf 88}, 257002 (2002).

\bibitem{ACENS02} A.~Abanov, A.V.~Chubukov, M.~Eschrig, M.R.~Norman, and
J.~Schmalian, Phys.~Rev.~Lett. {\bf 89}, 177002 (2002).

\bibitem{SA64} D.J.~Scalapino and P.W.~Anderson, Phys.~Rev. {\bf 33},
A921 (1964).

\bibitem{note1} In the superconducting state the value of $Z_1(\Delta_0)$ is
less than the normal state renormalization factor $1+\lambda = 1.5$ due to
the additional energy associated with the gap. Traditionally, when
$\Omega_0$ was large compared to $\Delta_0$, this effect was negligible.
Here, however, since $\Omega_0=1.5 \Delta_0$, the effective $\lambda$ for
the $s$-wave case is of order $2|g|^2 N(0)/(\Omega_0 + \Delta_0) =
0.6\lambda=0.3$. For the $d$-wave case, the effective $\lambda$ is also
reduced but not quite as much because of the nodes. In the following, we
will use $Z_1 (k, \omega=0)$ at the nodal point to give the nodal Fermi
velocity renormalization for the $d$-wave case.

\bibitem{Sca69} D.J.~Scalapino, in {\sl Superconductivity, Vol 1},
ed.\, by R.D.~Parks, Marcell Dekker, New York (1969).

\bibitem{Hof02} J.E.~Hoffman {\it et al.}, Science {\bf 295}, 466
(2002).

\bibitem{note2} We will address the important question of whether the
magnetic resonance mode is in fact sufficiently strongly coupled to the
quasi-particles to give the observed structure in a later paper. Different
conclusions regarding this have been reached in Refs
\onlinecite{KKA02} and \onlinecite{ACENS02}.

\bibitem{ASJL96} O.K.~Andersen, S.Y.~Savrasov, O.~Jepsen, and
A.I.~Liechtenstein, J.~Low Temp.~Solids {\bf 105}, 285 (1996). 

\bibitem{JADS98} O.~Jepsen, O.K.~Andersen, I.~Dasgupta, and S.~Savrasov, 
J.~Low Temp.~Solids {\bf 59}, 1718 (1998).

\bibitem{Bor02} S.V.~Borisenko {\it et al.}, cond-mat/0209435.

\bibitem{Kim03} T.K.~Kim {\it et al.}, cond-mat/0303422.

\bibitem{SS97} Z.-X.~Shen and J.R.~Schrieffer, Phys.~Rev.~Lett. {\bf
78}, 1771 (1997).

\bibitem{Nor97} M.R.~Norman {\it et al.}, Phys.~Rev.~Lett. {\bf 79}, 
3506 (1997).

\bibitem{ND98} M.R.~Norman and H.~Ding, Phys.~Rev.~B {\bf 57},
R11089 (1998).

\bibitem{EN02} M.~Eschrig and M.R.~Norman, Phys.~Rev.~Lett. {\bf 89},
277005 (2002).

\bibitem{NSB99} T.S.~Nunner, J.~Schmalian, and K.H.~Bennemann,
Phys.~Rev.~B {\bf 59}, 8859 (1999).

\end{thebibliography}
\end{document}